\documentclass[12pt]{article}
\usepackage{amsmath}
\usepackage[dvips]{graphicx}
\newlength{\extraspace}
\newlength{\extraspaces}
\catcode`\@=11
\def\numberbysection{\@addtoreset{equation}{section}
\def\theequation{\arabic{section}.\arabic{equation}}}
\newcommand{\newsection}[1]{
\vspace{7mm}
\pagebreak[3]
\addtocounter{section}{1}
\setcounter{equation}{0}
\setcounter{subsection}{0}
\setcounter{footnote}{0}
\begin{center}
{\large \textbf{\thesection. #1}}
\end{center}
\nopagebreak
\medskip
\nopagebreak
\hspace{3mm}}
\newcommand{\nonu}{\nonumber \\[.5mm]}
\newcommand{\A}{&\!\!\!}

\begin{document}
\addtolength{\baselineskip}{.7mm}
\thispagestyle{empty}
\begin{flushright}
STUPP--06--186 \\
\texttt{hep-th/0608198} \\ 
August, 2006
\end{flushright}
\vspace{20mm}
\begin{center}
{\Large \textbf{Holographic Chiral Phase Transition \\[2mm]
with Chemical Potential
}} \\[20mm]
\textsc{Norio Horigome}\footnote{
\texttt{e-mail: horigome@krishna.th.phy.saitama-u.ac.jp}} 
\hspace{1mm} and \hspace{2mm}
\textsc{Yoshiaki Tanii}\footnote{
\texttt{e-mail: tanii@phy.saitama-u.ac.jp}} \\[7mm]
\textit{Division of Material Science \\ 
Graduate School of Science and Engineering \\
Saitama University, Saitama 338-8570, Japan} \\[20mm]
\textbf{Abstract}\\[7mm]
{\parbox{13cm}{\hspace{5mm}
We discuss the Sakai-Sugimoto model at finite temperature and 
finite chemical potential. 
It is a holographic model of large $N_c$ QCD with $N_f$ massless 
quarks based on a D4/D8-$\overline{\text{D8}}$ brane system. 
The near horizon limit of the D4-branes and the probe approximation 
of the D8-$\overline{\text{D8}}$ pairs allow us to treat the 
D4-branes as a gravitational background and the D8-$\overline{\text{D8}}$ 
pairs as a probe which does not affect the background. 
We propose that the asymptotic value of a U(1) gauge field 
on the D8-$\overline{\text{D8}}$-branes is identified with the 
chemical potential for the baryon number. 
Using this chemical potential we analyze the phase structure of 
this model and find a chiral symmetry phase transition of the 
first order. 
}}
\end{center}
\vfill
\newpage
\setcounter{section}{0}
\setcounter{equation}{0}
\numberbysection
%
%
\newsection{Introduction}
The AdS/CFT correspondence 
\cite{Maldacena:1997re,Gubser:1998bc,Witten:1998qj} 
(see \cite{Aharony:1999ti} for a review) is a useful duality
between a string theory in ($d+1$)-dimensional 
anti de Sitter spacetime (times a compact space) 
and a $d$-dimensional conformal field theory. 
The AdS/CFT correspondence can be extended to more general 
cases of the string/gauge duality for non-conformal and 
non-supersymmetric theories. 
In this scheme one can discuss some features of the low energy QCD 
such as confinement and spontaneous chiral symmetry breaking. 
Such an approach to low energy behaviors of QCD in terms of the 
string/gauge duality is often called the holographic QCD 
\cite{Karch:2002sh,Sakai:2003wu,Babington:2003vm,
Kruczenski:2003uq,Ghoroku:2004sp,Antonyan:2006vw} 
(and references therein). 
\par
\begin{figure}[!t]
\begin{center}
\includegraphics{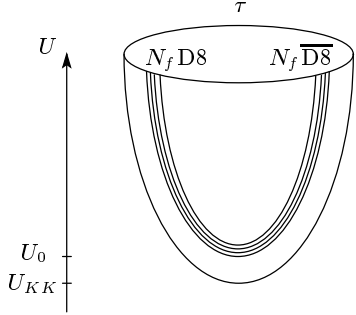}\\
\vspace{1mm}
\caption{A D8-$\overline{\text{D8}}$-brane configuration 
in the low temperature phase.}
\label{chiral-low}
\end{center}
\end{figure}

One of the interesting recent developments in the holographic 
approach to QCD is the D4/D8-$\overline{\text{D8}}$ model proposed 
by Sakai and Sugimoto \cite{Sakai:2004cn,Sakai:2005yt}. 
This brane system consists of $N_c$ D4-branes compactified on 
S${}^1$ and probe $N_f$ D8-$\overline{\text{D8}}$-brane pairs. 
The D4-branes are described by the extremal brane solution 
in the near horizon limit with one of the spatial directions 
along its world-volume compactified on S${}^1$. 
This gravitational background is dual to a five-dimensional 
gauge theory, which looks four-dimensional at energy scale below 
the compactification scale. 
This description is valid for the case 
$1 \ll g_{YM}^2 N_c \ll 1/g_{YM}^4$, 
where $g_{YM}$ is the four-dimensional gauge coupling. 
Imposing periodic boundary conditions on the bosons and 
anti-periodic ones on the fermions along the compactified 
direction, supersymmetry is explicitly broken.
The scalars and the fermions on the D4-branes become massive 
and are decoupled from the system at low energy. 
Thus one obtains a U($N_c$) pure gauge theory. 
To describe quarks in the fundamental representation of the gauge 
group U($N_c$) one introduces $N_f$ D8-$\overline{\text{D8}}$ pairs 
into the D4 background. 
The probe approximation $N_f \ll N_c$ \cite{Karch:2002sh} allows us 
to treat the $N_f$ D8-$\overline{\text{D8}}$ pairs as a probe, which 
does not affect the D4 background. A string connecting the D4-branes 
and the D8-branes ($\overline{\text{D8}}$-branes) represents 
a massless left-(right-)handed quark with $N_f$ flavors. 
Therefore one obtains a four-dimensional U($N_c$) gauge theory 
with $N_f$ flavored massless quarks in the fundamental 
representation of the gauge group at low energy. 
The $\text{U}(N_f)_L \times \text{U}(N_f)_R$ symmetry of 
the D8 and $\overline{\text{D8}}$-branes represents a chiral 
symmetry of the quarks. It was shown in ref.~\cite{Sakai:2004cn} 
that this chiral symmetry is broken to a diagonal symmetry 
U($N_f$)${}_V$ because of a configuration of the probe 
D8-$\overline{\text{D8}}$ pairs. 
By the topology of the D4-brane geometry they must have a curved 
configuration in which the D8-branes and the 
$\overline{\text{D8}}$-branes are connected each other 
(see Fig.~\ref{chiral-low}), which breaks the chiral symmetry. 
\par
\begin{figure}[!t]
\begin{center}
\includegraphics{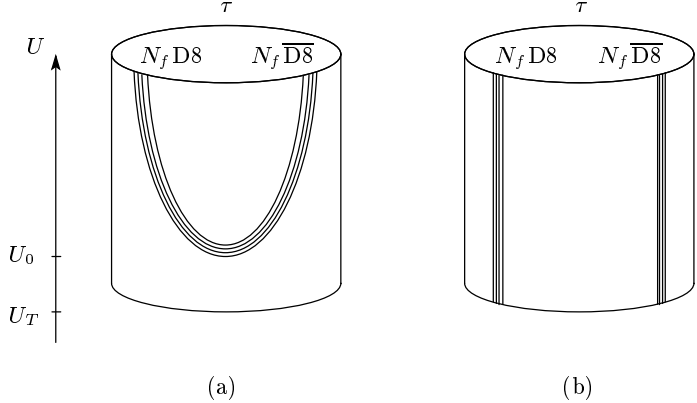}
\caption{D8-$\overline{\text{D8}}$-brane configurations 
in the high temperature phase.}
\label{chiral-high}
\end{center}
\end{figure}

The Sakai-Sugimoto model was also studied at finite temperature 
$T$ \cite{Aharony:2006da,Parnachev:2006dn,Peeters:2006iu}. 
There are two D4-brane geometries which represent a low 
temperature phase and a high temperature phase respectively. 
A phase transition between 
these phases occurs at a certain critical temperature $T_c$. 
This transition corresponds to a confinement/deconfinement 
transition in a dual gauge theory \cite{Witten:1998zw}. 
For each of the phases one can introduce 
probe D8-$\overline{\text{D8}}$ pairs. 
The only configuration of the D8-$\overline{\text{D8}}$ pairs 
which can realize in the low temperature phase $T < T_c$ 
is a curved D8-$\overline{\text{D8}}$ configuration as 
in the zero temperature case. Thus the chiral symmetry is 
always broken in the low temperature phase. 
On the other hand, another configuration exists in addition 
to the curved one in the high temperature phase $T > T_c$. 
The new configuration consists of straight D8-branes 
and $\overline{\text{D8}}$-branes 
disconnected each other (see Fig.~\ref{chiral-high}). 
The chiral symmetry is unbroken for this configuration. 
A chiral symmetry phase transition can occur between these 
two configurations at a certain temperature 
\cite{Aharony:2006da,Parnachev:2006dn}.
\par
The purpose of the present paper is to analyze the Sakai-Sugimoto 
model at finite temperature $T$ and finite baryon number chemical 
potential $\mu$. We introduce a non-vanishing background U(1) gauge 
field on the probe D8-$\overline{\text{D8}}$-brane world-volume. 
The asymptotic value of this gauge field background is related to 
the baryon number chemical potential. 
There are several related works in which chemical potentials 
are introduced as asymptotic values of gauge fields 
\cite{Chamblin:1999tk,Cvetic:1999ne,Chamblin:1999hg,Albash:2006bs,Apreda:2005yz}.
The gauge fields and the chemical potentials considered there, 
however, are those for the R symmetry in the bulk geometry
%
%
or for the isospin symmetry on the probe brane world-volume. 
In contrast, we consider the chemical potential 
for the baryon number U(1)${}_V$ symmetry on the probe brane. 
The D8-$\overline{\text{D8}}$-brane configuration and the gauge 
field background are determined by equations derived from 
the Dirac-Born-Infeld effective action on the world-volume 
of the probe branes. 
By solving these equations and comparing values of the effective 
action for the solutions we discuss a chiral symmetry breaking 
as in refs.~\cite{Sakai:2004cn,Aharony:2006da,Parnachev:2006dn}. 
\par
\begin{figure}[!t]
\begin{center}
\includegraphics{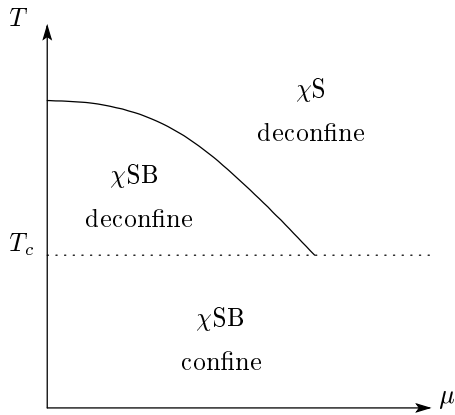}\\
\vspace{3mm}
\caption{The phase diagram of the dual gauge theory.}
\label{phase}
\end{center}
\end{figure}

We can summarize our results as follows. 
In the low temperature phase $T < T_c$ of the 
confinement/deconfinement transition there is a unique solution 
for the probe brane configuration and the gauge field background. 
As in the case without the gauge field background the probe 
branes have a curved configuration and the chiral symmetry is 
always broken. In the high temperature phase $T > T_c$ 
there are two types of solutions. 
One solution has a curved brane configuration and the chiral 
symmetry is broken. The other solution has a straight brane 
configuration and the chiral symmetry is unbroken. 
A phase transition between these two solutions occurs 
at certain temperature and chemical potential. 
The transition is of the first order. 
There is a critical value of the chemical potential above which 
the phase transition never occurs for any temperature. 
A qualitative feature of the phase diagram is shown 
in Fig.~\ref{phase}. 
\par
The organization of this paper is as follows. 
In section 2 we review the bulk geometry of the Sakai-Sugimoto 
model at finite temperature and set up the effective action for 
the probe brane configuration and the U(1) gauge field. 
In sections 3 and 4 we obtain solutions for the brane configuration 
and the gauge filed and discuss the chiral symmetry breaking 
in the low and high temperature phases respectively. 
Section 5 is devoted to some discussions. 
\par
While preparing the manuscript of this paper, we have received 
a paper \cite{Kim:2006gp}, in which a gauge field background 
on the probe branes is used to represent the baryon number 
chemical potential in studying hadronic matters in the 
Sakai-Sugimoto model.  
After submitting the manuscript of this paper for publication, 
we have received a paper \cite{Parnachev:2006ev} discussing 
the chiral phase transition in the D4/D8-$\overline{\text{D8}}$ 
model, in which an error in the first version of our paper 
was pointed out. 
%
%
\newsection{D4/D8-$\overline{\text{D8}}$ brane system}
The Sakai-Sugimoto model \cite{Sakai:2004cn,Sakai:2005yt} 
is based on a D4/D8-$\overline{\text{D8}}$ brane system 
consisting of S${}^1$ compactified $N_c$ D4-branes and $N_f$ 
D8-$\overline{\text{D8}}$-brane pairs transverse to 
the S${}^1$. The brane configuration of the system is 
\begin{equation}
\begin{tabular}{ccccccccccc}
& $t$ & $x^1$ & $x^2$ & $x^3$ & $\tau$ & $U$ 
& $\theta^1$ & $\theta^2$ & $\theta^3$ & $\theta^4$ \\ \hline
D4 & $\circ$ & $\circ$ & $\circ$ & $\circ$ & $\circ$ & $-$ 
& $-$ & $-$ & $-$ & $-$ \\
D8-$\overline{\text{D8}}$ & $\circ$ & $\circ$ & $\circ$ & $\circ$ 
& $-$ & $\circ$ & $\circ$ & $\circ$ & $\circ$ & $\circ$ \\
\end{tabular}
\end{equation}
with $\tau$ and $\theta$'s being coordinates of S${}^1$ and 
S${}^4$ respectively. The period of $\tau$ is denoted as 
$\delta\tau = 2\pi/M_{KK}$. 
In the large $N_c$ limit and the near horizon limit the D4-branes 
are described by a bulk background geometry, which is a classical 
solution of the type IIA supergravity in ten dimensions. 
Assuming $N_f \ll N_c$ the D8-$\overline{\text{D8}}$ pairs are 
treated as a probe which does not affect the bulk background. 
\par
The finite temperature behavior of the Sakai-Sugimoto model was 
discussed in \cite{Aharony:2006da, Parnachev:2006dn,Peeters:2006iu}. 
The bulk background geometry is represented by a metric with 
a periodic Euclidean time coordinate 
$t_E \equiv i t \sim t_E + \delta t_E$ 
in addition to the periodic $\tau$. 
The period of $t_E$ is the inverse temperature $\delta t_E = 1/T$. 
There are two such Euclidean solutions which have an 
appropriate asymptotic boundary behavior. 
One of them is the Euclidean version of the extremal D4-brane 
geometry compactified on S${}^1$ with the metric 
\begin{eqnarray}
ds^2 \A = \A \left( \frac{U}{R} \right)^{\frac{3}{2}} 
       \left( dt_E^2 + \delta_{ij}dx^{i}dx^{j} + f(U) d\tau^2 \right)
       + \left( \frac{R}{U} \right)^{\frac{3}{2}} 
       \left( \frac{dU^2}{f(U)} + U^2 d\Omega^2_4 \right), \nonu
\A\A \qquad\quad f(U) = 1 - \frac{U_{KK}^3}{U^3}, \qquad
U_{KK} = \frac{4}{9} R^3 M_{KK}^2, 
\label{metric-l}
\end{eqnarray}
where $d\Omega_4^2$ is the metric of S${}^4$ and 
$R^3 = \pi g_s N_c l_s^3$ with $g_s$ and $l_s$ being 
the string coupling and the string length. 
The parameter $U_{KK}$ must be related to $M_{KK}$ as above 
to avoid a singularity of the metric at $U=U_{KK}$. 
With this relation the $\tau$-$U$ 
submanifold has a cigar-like form with a tip at $U=U_{KK}$. 
The dilaton $\phi$ and the RR 3-form $C_3$ are given by 
\begin{equation}
e^{\phi} = g_s \left( \frac{U}{R} \right)^{\frac{3}{4}}, \qquad 
F_4 = dC_3 = \frac{2 \pi N_c}{V_4}\epsilon_4, 
\label{dilaton}
\end{equation}
where $\epsilon_4$ and $V_4$ are the volume form and the volume 
of S${}^4$. 
The other solution is the Euclidean version of the non-extremal
D4-brane geometry 
\begin{eqnarray}
ds^2 \A = \A \left( \frac{U}{R} \right)^{\frac{3}{2}} 
       \left( \tilde{f}(U) dt_E^2 + \delta_{ij}dx^{i}dx^{j} 
       + d\tau^2 \right)
       + \left( \frac{R}{U} \right)^{\frac{3}{2}} 
       \left( \frac{dU^2}{\tilde{f}(U)} + U^2 d\Omega^2_4 \right), \nonu
\A\A \qquad\quad \tilde{f}(U) = 1- \frac{U_{T}^3}{U^3}, \qquad
U_T = \frac{16\pi^2}{9} R^3 T^2  
\label{metric-h}
\end{eqnarray}
with the dilaton and the RR 3-form given in eq.~(\ref{dilaton}). 
The parameter $U_T$ must be related to $T$ as above to avoid a 
singularity of the metric at $U=U_T$. The $t_E$-$U$ submanifold 
has a cigar-like form with a tip at $U=U_T$. 
It is obvious that these two backgrounds are related by 
interchanging $\tau$, $U_{KK}$ and $t_E$, $U_T$. 
\par
It was shown \cite{Witten:1998zw,Kruczenski:2003uq,Aharony:2006da} 
that the background (\ref{metric-l}) is dominant at low temperature, 
while (\ref{metric-h}) is dominant at high temperature 
by comparing values of the Euclidean supergravity action 
for these backgrounds. 
A phase transition between these backgrounds occurs at the 
temperature for $U_T = U_{KK}$, i.e. $T_c = M_{KK}/(2\pi)$. 
This phase transition is of the first order and represents 
a confinement/deconfinement transition \cite{Witten:1998zw}. 
\par
In refs.~\cite{Aharony:2006da,Parnachev:2006dn} $N_f$ 
D8-$\overline{\text{D8}}$ pairs were introduced as a probe 
in the backgrounds (\ref{metric-l}), (\ref{metric-h}).
The effective action of the D8-branes consists of the 
Dirac-Born-Infeld action and the Chern-Simons term
\begin{equation}
S_{\text{D8}} = T_8 \int d^9x \, e^{-\phi} \, \text{Tr} 
\sqrt{\det (g_{MN} + 2\pi\alpha'F_{MN})} 
- \frac{i}{48 \pi^3} \int C_3 \; \text{Tr} F^3, 
\label{dbics}
\end{equation}
where $g_{MN}$ and $F_{MN}=\partial_M A_N - \partial_N A_M 
-i \left[ A_M , A_N \right]$ ($M,N = 0,1,\cdots,8$) 
are the induced metric and the field strength of the U($N_f$) 
gauge field $A_M$ on the D8-branes. 
$T_8$ is the tension of the D8-brane and $\alpha'=l_s^2$ is 
the Regge slope parameter. 
The effective action for the $\overline{\text{D8}}$-branes has 
a similar form. The total effective action has a gauge symmetry 
\begin{equation}
\text{U}(N_f)_L \times \text{U}(N_f)_R 
= \text{SU}(N_f)_L \times \text{SU}(N_f)_R \times \text{U}(1)_V 
\times \text{U}(1)_A,  
\label{flavorsymmetry}
\end{equation}
where U($N_f$)${}_L$ and U($N_f$)${}_R$ are symmetries 
of $N_f$ D8 and $\overline{\text{D8}}$-branes respectively. 
It was argued in ref.~\cite{Sakai:2004cn} that this gauge symmetry 
corresponds to a flavor chiral symmetry of quarks. 
The total effective action can be written in the form 
(\ref{dbics}) with the integrations being over the whole of 
the D8-$\overline{\text{D8}}$ world-volume. 
We use this form of the effective action in the following. 
\par
In refs.~\cite{Sakai:2004cn,Aharony:2006da} the gauge fields 
on the probe branes are treated as 
fluctuations representing the hadron spectrum. 
In this paper we consider a background gauge field. 
We assume that only the Euclidean time component of the 
U(1) gauge field has a non-vanishing background. 
We will see that it corresponds to an introduction of the baryon 
number chemical potential. We use a physical gauge for D8-brane 
world-volume reparametrizations and use the spacetime 
coordinates other than $\tau$ as the world-volume coordinates. 
Then, D8 and $\overline{\text{D8}}$-brane configurations 
are determined by $\tau$ as a function of those world-volume 
coordinates. We make an ansatz that $A_0$ and $\tau$ depend 
only on the coordinate $U$
\begin{equation}
\tau = \tau(U), \qquad A_0 = A_0(U). 
\label{ansatz}
\end{equation}
By this ansatz the Chern-Simons term in eq.~(\ref{dbics}) 
vanishes and does not concern us. 
%
%
\newsection{Low temperature phase}
In the low temperature phase the geometry (\ref{metric-l}) is 
dominant. Using the ansatz (\ref{ansatz}) 
the induced metric $g_{MN}$ on the probe D8-branes is given by 
\begin{eqnarray}
ds^2 \A = \A \left( \frac{U}{R} \right)^{\frac{3}{2}}
\left( dt_E^2 + \delta_{ij}dx^idx^j \right) \nonu
\A\A + \left[ \left( \frac{U}{R} \right)^{\frac{3}{2}} f(U) (\tau'(U))^2 
+ \left( \frac{R}{U} \right)^{\frac{3}{2}} \frac{1}{f(U)} \right] dU^2 
+ \left( \frac{R}{U} \right)^{\frac{3}{2}} U^2d\Omega^2_4, 
\label{induced-metric-1}
\end{eqnarray}
where $\tau' = \frac{d\tau}{dU}$.
Then, the effective action of the D8-branes (\ref{dbics}) becomes
\begin{equation}
S_{\text{D8}} = 
\frac{N_f T_8 V_4}{g_s} \int d^4x 
\, dU U^4 
\left[ f \, (\tau')^2 + \left( \frac{R}{U} \right)^3 
\left( f^{-1} - \left( 2\pi\alpha' A_0' \right)^2 \right) 
\right]^{\frac{1}{2}}, 
\label{action-a}
\end{equation}
where $A_0' = \frac{dA_0}{dU}$. 
\par
This action leads to equations of motion for $\tau(U)$ and $A_0(U)$ 
\begin{eqnarray}
\frac{d}{dU} 
\left[ \frac{U^4 f \, \tau'}
{\sqrt{f \, (\tau')^2 + \left( \frac{R}{U}\right)^3 
\left( f^{-1} - \left( 2\pi\alpha' A_0' \right)^2 \right) }} 
\right] \A = \A 0, \nonu
\frac{d}{dU} 
\left[ \frac{U^4 \left( \frac{R}{U} \right)^3 A_0'} 
{ \sqrt{ f \left( \tau' \right)^2 + \left( \frac{R}{U} \right)^3 
\left( f^{-1} - \left( 2\pi\alpha' A_0' \right)^2 \right)}
} \right] \A = \A 0,  
\label{eom-low}
\end{eqnarray}
which can be easily integrated once. We obtain 
\begin{eqnarray}
(\tau'(U))^2 \A = \A \frac{
\left( U_0^8 + C^2 \left( \frac{U_0}{R} \right)^3 \right) f(U_0) 
\left( \frac{R}{U} \right)^6 }
{f(U)^2 \left[ 
\left( \frac{R}{U} \right)^3 \left( U^8 f(U) - U_0^8 f(U_0) \right) 
+ C^2 \left( f(U) - f(U_0) \left( \frac{U_0}{U} \right)^3 \right) 
\right] }, \nonu
\left( 2\pi\alpha' A_0'(U) \right)^2 
\A = \A \frac{C^2}
{ \left( \frac{R}{U} \right)^3 \left( U^8 f(U) - U_0^8 f(U_0) \right) 
+ C^2 \left( f(U) - f(U_0) \left( \frac{U_0}{U} \right)^3 \right) }, 
\label{solution-low}
\end{eqnarray}
where $C$ and $U_0$ are integration constants. 
As in the zero temperature case \cite{Sakai:2004cn} 
and the low temperature phase in ref.~\cite{Aharony:2006da} 
we have imposed a condition $\tau'(U_0) = \infty$. 
A typical configuration of $\tau(U)$ is shown in 
Fig.~\ref{chiral-low}. Since there is no place for the D8 and 
$\overline{\text{D8}}$-branes to end, they are connected 
at $U=U_0$. 
%
%
We also impose the boundary condition $A_{0}(\infty) = \mu$ 
at the both ends of the D8-$\overline{\text{D8}}$ world-volume, 
where $\mu$ is a constant. We will identify this constant with 
the chemical potential for the baryon number later. 
Solving eq.~(\ref{solution-low}) with this boundary condition 
we find for $U \sim U_0$ 
\begin{equation}
A_0(U) \sim A_0(U_0) + {\rm const.} \times C \, |U-U_0|^{\frac{1}{2}}. 
\end{equation}
This solution is singular at $U=U_0$ and does not actually 
satisfy the original equation (\ref{eom-low}) unless $C=0$. 
Therefore, we must choose $C=0$ and obtain $A_0(U) = \mu$. 
%
%
Because of the connected configuration of the D8 and 
$\overline{\text{D8}}$ branes 
the chiral symmetry $\text{U}(N_f)_L \times \text{U}(N_f)_R$ 
on the probe D8-$\overline{\text{D8}}$ pairs is always broken 
to a diagonal subgroup $\text{U}(N_f)_V$ in the low temperature 
phase. The situation is the same as in the cases without the gauge 
field background \cite{Sakai:2004cn, Aharony:2006da}. 
\par
Instead of using the constant $U_0$ to parametrize the solution 
we can also use the $U=\infty$ asymptotic separation $L$ between 
the D8 and $\overline{\text{D8}}$-branes in the $\tau$-direction. 
It is related to $U_0$ by 
\begin{equation}
L = 2 \int_{U_0}^\infty dU \, \tau'(U), 
\label{asymptoticL}
\end{equation}
where $\tau'(U)$ is given in eq.~(\ref{solution-low}) with $C=0$. 
\par
\par
Substituting eq.~(\ref{solution-low}) with $C=0$ into the 
action (\ref{action-a}) and introducing new variables 
$u = U/U_0$, $u_{KK} = U_{KK}/U_0$ 
and $f(u)=1-u_{KK}^3/u^3$ the effective action becomes
\begin{equation}
S_{\text{D8}} 
= \tilde{T}_8 \int_{1}^{\infty} du \, u^4
\sqrt{ \frac{u^5}{u^{8} f(u) - f(1)}}, 
\end{equation}
where 
\begin{equation}
\tilde{T}_8 = \frac{N_f T_8V_4}{g_s} 
\left( R^3 U_0^7 \right)^{\frac{1}{2}} \int d^4x.
\label{ct8}
\end{equation}
Note that this reproduce the result 
in ref.~\cite{Aharony:2006da}.


%
\newsection{High temperature phase}
In the high temperature phase the geometry (\ref{metric-h}) 
is dominant. Using the ansatz (\ref{ansatz}) the induced metric 
on the probe D8-branes is 
\begin{eqnarray}
ds^2 \A = \A \left( \frac{U}{R} \right)^{\frac{3}{2}} 
\left( \tilde{f}(U) dt_E^2 + \delta_{ij}dx^idx^j \right) \nonu
\A\A + \left[ \left( \frac{U}{R} \right)^{\frac{3}{2}} (\tau'(U))^2 
+ \left( \frac{R}{U} \right)^{\frac{3}{2}} \frac{1}{\tilde{f}(U)} 
\right] dU^2 
+ \left( \frac{R}{U} \right)^{\frac{3}{2}} U^2 d\Omega^2_4
\label{metric-high}
\end{eqnarray}
and the effective action of the D8-branes (\ref{dbics}) becomes
\begin{equation}
S_{\text{D8}} 
= \frac{N_f T_8V_4}{g_s} \int d^4x \, dU U^4 
  \left[ \tilde{f} \, (\tau')^2 
  + \left( \frac{R}{U} \right)^3 
  \left( 1 - \left( 2\pi \alpha' A_0' \right)^2 \right) 
  \right]^{\frac{1}{2}}. 
\label{action-b}
\end{equation}
This action leads to equations of motion for $\tau(U)$ and $A_0(U)$ 
\begin{eqnarray}
\frac{d}{dU} 
\left[ \frac{ U^4 \tilde{f} \, \tau' }
{ \sqrt{\tilde{f} \, (\tau')^2 
+ \left( \frac{R}{U} \right)^3 
\left( 1 - \left( 2\pi \alpha' A_0' \right)^2 \right)}} 
\right] \A = \A 0, \nonu
\frac{d}{dU}
\left[ \frac{U^4 \left( \frac{R}{U} \right)^3 A_0'}
{\sqrt{ \tilde{f} \, (\tau')^2 
+ \left( \frac{R}{U} \right)^3 
\left( 1 - \left( 2\pi \alpha' A_0' \right)^2 \right) }} 
\right] \A = \A 0,  
\label{eom-high}
\end{eqnarray}
which can be easily integrated once as before. 
As in the case without the gauge field 
\cite{Aharony:2006da, Parnachev:2006dn} there are two types 
of solutions in the high temperature phase. 
\par
One solution is similar to the one in the low temperature phase. 
The integration of eq.~(\ref{eom-high}) gives 
\begin{equation}
\left( \tau' (U) \right)^{2} = \frac{U_{0}^{8} \tilde{f}(U_{0})}
{\left( \frac{U}{R} \right)^{3} \tilde{f}(U)
\left( U^{8} \tilde{f}(U) - U_{0}^{8} \tilde{f}(U_{0}) \right)}, 
\qquad A_{0}(U) = \mu.
\label{a0-curved}
\end{equation}
where $U_0$ is an integration constant.
As before we have imposed the boundary conditions
$\tau'(U_0) = \infty$ and $A_0(\infty)=\mu$. 
A typical configuration of $\tau(U)$ is shown in 
Fig.~\ref{chiral-high} (a). 
The chiral symmetry $\text{U}(N_f)_L \times \text{U}(N_f)_R$ 
is broken to a diagonal subgroup $\text{U}(N_f)_V$. 
Substituting eq.~(\ref{a0-curved}) into eq.~(\ref{action-b}) 
the effective action becomes 
\begin{equation}
S_{\text{D8}}^{\text{U}} 
= \tilde{T}_8 \int_{1}^{\infty} du \ u^{5}
\sqrt{ \frac{u^3 \tilde{f}(u)}
{ u^{8} \tilde{f}(u) - \tilde{f}(1)}}, 
\label{action-U}
\end{equation}
where we have rescaled the variables as $u = U/U_0$, 
$u_T = U_T/U_0$, $\tilde{f}(u)=1-u_T^3/u^3$, and $\tilde{T}_8$ 
is given in eq.~(\ref{ct8}). 
\par
Instead of using $U_0$ we can also use the asymptotic separation $L$ 
in eq.~(\ref{asymptoticL}) to parametrize the solution, which is 
more convenient when comparing this solution to the other one. 
The relation between $L$ and $U_0$ is obtained from 
eqs.~(\ref{asymptoticL}) and (\ref{a0-curved}) as 
\begin{equation}
L = \left( \frac{R^3}{U_0} \right)^{\frac{1}{2}} F(u_T), 
\label{separationL}
\end{equation}
where 
\begin{equation}
F(u_T) 
= 2 \int_1^\infty du \,
{\sqrt\frac{\tilde{f}(1)}{u^3 \tilde{f}(u)
\left( u^8 \tilde{f}(u) - \tilde{f}(1) \right)}}. 
\end{equation}
\par
For the other solution the first integration of 
eq.~(\ref{eom-high}) gives 
\begin{equation}
\tau'(U) = 0, \qquad
\left( 2\pi\alpha'A_0'(U) \right)^2 
= \frac{C^2}{U^8 \left( \frac{R}{U} \right)^3 + C^2}, 
\label{a0-straight}
\end{equation}
where $C$ is an integration constant. 
%
%
%
$\tau'(U) = 0$ is the trivial solution of (\ref{eom-high}). 
A typical configuration is shown in Fig.~\ref{chiral-high} (b). 
It describes a situation that 
the probe D8 and $\overline{\text{D8}}$-branes separately 
extend along the $U$-direction in straight lines. 
The separation between the D8 and $\overline{\text{D8}}$-branes 
is chosen to be the same as the asymptotic separation $L$ 
in the previous solution. 
The chiral symmetry $\text{U}(N_f)_L \times \text{U}(N_f)_R$ 
is unbroken in this case. 
Substituting eq.~(\ref{a0-straight}) into eq.~(\ref{action-b})
and using the rescaled variables as in eq.~(\ref{action-U}) 
the effective action becomes 
\begin{equation}
S_{\text{D8}}^{||} 
= \tilde{T}_8 \int_{u_T}^{\infty} du \, \frac{u^5}{\sqrt{u^5 + c^2}},
\label{action-||}
\end{equation}
where
\begin{equation}
c^2 = \frac{C^2}{R^3 U_0^5}.
\end{equation}
\par
To determine which of the two solutions is dominant we compare 
the values of the effective action. From eqs.~(\ref{action-U}), 
(\ref{action-||}) we obtain the difference as 
\begin{eqnarray}
\Delta S 
\A \equiv \A \frac{S_{\text{D8}}^{\text{U}} 
- S_{\text{D8}}^{||}}{\tilde{T}_8} \nonu
\A = \A \int_{1}^{\infty} du \ u^{5} 
\left[ \sqrt{ \frac{u^3 \tilde{f}(u)}
{ u^{8} \tilde{f}(u) - \tilde{f}(1) }}
-  \frac{1}{ \sqrt{ u^{5} + c^2 } } \right]
- \int_{u_T}^{1} du \; 
\frac{u^{5}}{ \sqrt{ u^{5} + c^2 } }. 
\label{delta-s}
\end{eqnarray}
\begin{figure}[!t]
\begin{center}
\includegraphics{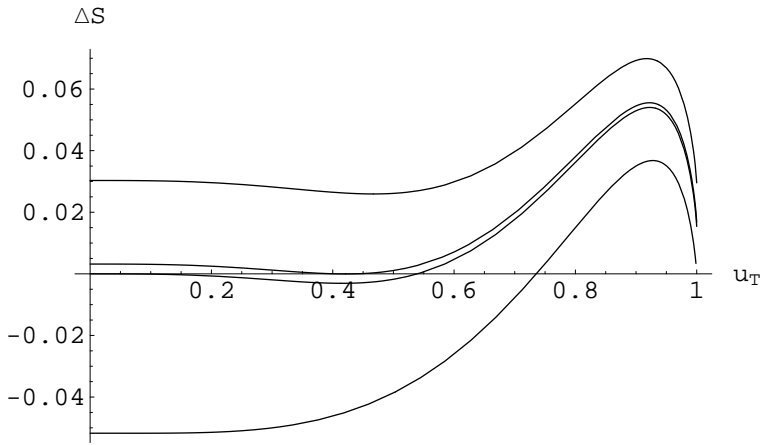}
\caption{$\Delta S$ as a function of $u_T$ for various values of $c$. 
{}From the bottom to the top each line represents the case for 
$c =$ 0, 0.2158, 0.2252, 0.3 respectively.}
\label{graph-delta-s}
\end{center}
\end{figure}
\par
For $\Delta S < 0$ the curved configuration (\ref{a0-curved}) 
is dominant and the chiral symmetry is broken, while for 
$\Delta S > 0$ the straight configuration (\ref{a0-straight}) 
is dominant and the chiral symmetry is unbroken. 
Although the integrals in eqs.\ (\ref{action-U}), 
(\ref{action-||}) are divergent at $U=\infty$, 
the difference is finite due to the same asymptotic behaviors 
of $\tau(U)$ and $A_0(U)$. 
We evaluate eq.~(\ref{delta-s}) by numerical calculations. 
For that purpose it is more convenient 
to change an integration variable to $z=u^{-3}$, which has a 
finite interval $0 \leq z \leq 1$ for $1 \leq u < \infty$. 
The result of the calculations is shown in 
Fig.~\ref{graph-delta-s}. 
The behaviors of $\Delta S$ 
as a function of $u_T$ for various values of $c$ are given. 
The special case $c=0$ reduces to the result in 
ref.~\cite{Aharony:2006da}. 
In this case $\Delta S$ is positive for $u_T$ larger 
than a certain value $u_{T0}$ and negative for $u_T < u_{T0}$. 
The chiral symmetry is broken for $u_T < u_{T0}$ and unbroken 
for $u_T > u_{T0}$. 
The point $u_T = u_{T0}$ is a phase transition point. 
This phase transition is of the first order since two different 
configurations in Fig.~\ref{chiral-high} are possible 
at the transition point. 
As $c$ increases, the transition point $u_{T0}$ decreases. 
When $c > 0.2158$, there appears a new region near $u_T = 0$ 
in which $\Delta S > 0$. 
When $c > 0.2252$, $\Delta S$ is positive for all values 
of $u_T$ and the chiral symmetry is always unbroken. 
\par
\begin{figure}[!t]
\begin{center}
\begin{tabular}{cc}
\includegraphics[width=10cm]{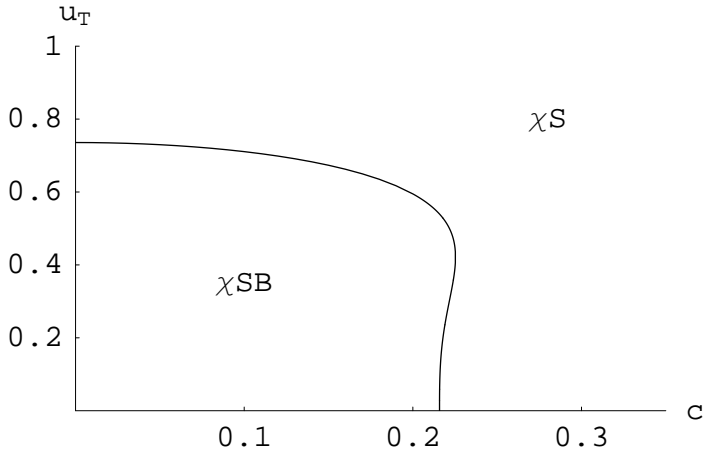}
\end{tabular}
\caption{The phase diagram in the $c$-$u_T$ space.}
\label{phase-ut}
\end{center}
\end{figure}

{}From these results we can draw a phase diagram in the $c$-$u_T$ 
space as shown in Fig.~\ref{phase-ut}. 
The chiral symmetry is broken in the region of small $c$ and 
small $u_T$ and unbroken outside of it. 
Note that we are considering here the high temperature phase 
of the confinement/deconfinement transition and only the part 
$u_T > u_{KK}$ of this diagram is valid. 
\par
It is more appropriate, however, to draw it in the space 
of the temperature $T$ and the baryon number chemical potential $\mu$. 
{}From eq.~(\ref{metric-h}) the temperature $T$ is related to 
$u_T$ as 
\begin{equation}
T = \frac{3}{4\pi} \left( \frac{U_0}{R^3} \right)^{\frac{1}{2}} 
\sqrt{u_T}
= \frac{3}{4\pi} \frac{\sqrt{u_T}}{L}  F(u_T), 
\label{temperature}
\end{equation}
where we have used eq.~(\ref{separationL}). 
\par
The relation of the chemical potential $\mu$ to $u_T$ and $c$ can be 
obtained as follows. From eq.~(\ref{a0-straight}) the large $U$ 
behavior of $A_0(U)$ has a form 
\begin{equation}
A_0(U) \sim \mu + \frac{v}{U^{\frac{3}{2}}}, 
\label{a-behavior}
\end{equation}
where $\mu$ and $v$ are constants. 
We have chosen the same value $\mu$ for the constant term 
as in the curved solution (\ref{a0-curved}). 
According to the AdS/CFT dictionary \cite{Aharony:1999ti} 
for a massless vector field in a six-dimensional bulk, 
$\mu$ is a source coupled to an operator of dimension four 
${\cal O}_{4}$ on a five-dimensional boundary. 
The U(1) gauge field $A_0$ defined on the whole of the 
D8-$\overline{\text{D8}}$ world-volume contains 
the gauge fields for both of U(1)${}_V$ and U(1)${}_A$ 
in the flavor symmetry (\ref{flavorsymmetry}). 
The part of $A_0$ which is symmetric for an interchange of D8 
and $\overline{\text{D8}}$ corresponds to U(1)${}_V$, while 
the part which is antisymmetric corresponds to U(1)${}_A$ 
\cite{Sakai:2004cn,Kim:2006gp}. 
Since the constant term $\mu$ is symmetric, 
it is a background value of the U(1)${}_V$ gauge field coupled 
to the baryon number density ${\cal O}_4$, and $\mu$ is 
the baryon number chemical potential. 
\par
Integration of eq.~(\ref{a0-straight}) determines $A_0(U)$ up 
to a constant term ($\mu$ in eq.~(\ref{a-behavior})). 
We can fix this constant term as follows. 
We first require that $A_0(U)$ vanishes at $U=U_T$ because 
of the regularity. 
%
%
To see this we first change the coordinates from $(U, t_E)$ to 
$(r, \theta)$ defined by 
\begin{equation}
U^{3} = U_{T}^{3} + U_{T} r^{2}, \qquad
\theta = \frac{3}{2} \left( \frac{U_{T}}{R^{3}} 
\right)^{\frac{1}{2}} t_{E}. 
\end{equation}
{}From the induced metric (\ref{metric-high}) with $\tau'(U) = 0$ 
we see that $(r, \theta)$ are the polar coordinates near 
the point $U=U_T$. The point $U=U_T$ corresponds to the origin $r=0$ 
and should be treated with care since the polar coordinates 
are not good coordinates near the origin. 
It is better to use the Cartesian coordinates 
\begin{equation}
y = r \cos \theta, \qquad 
z = r \sin \theta.
\end{equation}
The relation between $A_0$ and the components $A_y, A_z$ in the 
coordinates $(y, z)$ is obtained from 
$A_{0} dt_{E} = A_{y} dy + A_{z} dz$ as 
\begin{equation}
A_{0} = \frac{3}{2} \left( \frac{U_{T}}{R^{3}} \right)^{\frac{1}{2}} r
		\left( - A_{y} \sin \theta + A_{z} \cos \theta \right).
\end{equation}
Since we require that $A_{y}$ and $A_{z}$ are regular at the origin 
$r=0$, $A_{0}(U)$ must vanish at $U=U_{T}$. 
We also note that although $A_0(U)$ is a gauge dependent quantity, 
it must vanish at $U=U_T$ in any gauge. Only the gauge transformations 
which preserve the condition $A_0(U_T)=0$ are allowed. 
\par
The vanishing of $A_0(U)$ at $U=U_T$ fixes the constant term 
in this case and we find 
\begin{equation}
A_0(U) = \frac{U_0}{2\pi\alpha'} 
\int_{u_T}^{u} du' \, \sqrt{ \frac{c^2}{ u^{'5} + c^2 }}. 
\end{equation}
The chemical potential $\mu$ is obtained as the asymptotic value 
for $U = \infty$ 
\begin{equation}
\mu = A_0(\infty) = \frac{R^3}{2\pi \alpha'L^2} \, 
( F(u_T) )^2 
\int_{u_T}^{\infty} du \, \sqrt{ \frac{c^2}{ u^5 + c^2 }},  
\label{a0infty-straight}
\end{equation}
where we have used eq.~(\ref{separationL}) to eliminate $U_0$. 
This gives an expression of the chemical potential in terms of 
$u_T$ and $c$. 
\par
\begin{figure}[!t]
\begin{center}
\begin{tabular}{cc}
\includegraphics[width=9cm]{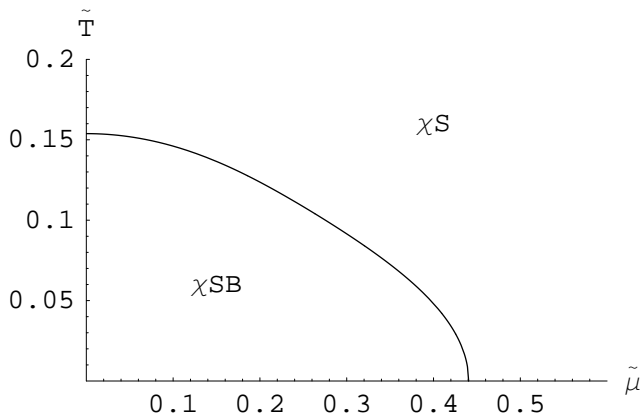} 
\end{tabular}
\caption{The phase diagram in the $\tilde{\mu}$-$\tilde{T}$ space.}
\label{phase-T}
\end{center}
\end{figure}

Using eqs.~(\ref{temperature}), (\ref{a0infty-straight}) 
we can convert the phase diagram in Fig.~\ref{phase-ut} 
to that in the $\mu$-$T$ space by numerical calculations. 
Using dimensionless variables 
\begin{equation}
\tilde{T} = L T, \qquad 
\tilde{\mu} = \frac{2\pi\alpha'L^2}{R^3}\mu. 
\label{dimensionless}
\end{equation}
the phase diagram in the $\tilde{\mu}$-$\tilde{T}$ space 
is shown in Fig.~\ref{phase-T}. 
Only the part of this diagram for the high 
temperature phase of the confinement/deconfinement transition, 
i.e. $\tilde{T} > \tilde{T}_c = L M_{KK}/(2\pi)$ is valid. 
Therefore, our result of the phase diagram looks like 
Fig.~\ref{phase} as we explained in Introduction. 
The orders of $T$ and $\mu$ at the transition points can be 
estimated from eq.~(\ref{dimensionless}). 
Using $R^3 = g_{YM}^2 N_c l_s^2/(2M_{KK})$ and 
$\tilde{\mu}, \tilde{T} = {\cal O}(1)$ we obtain 
\begin{equation}
T = {\cal O}(L^{-1}), \qquad
\mu = {\cal O}(g_{YM}^2 N_c L^{-2} M_{KK}^{-1}). 
\end{equation}
If we assume $L = {\cal O}(M_{KK}^{-1})$, the transition 
temperature is of the order of the compactification scale 
$M_{KK}$, and the chemical potential is of the order 
$g_{YM}^2 N_c M_{KK}$, which is much larger than $M_{KK}$ 
since $g_{YM}^2 N_c \gg 1$. 
%
%
\newsection{Discussions}
We analyzed the Sakai-Sugimoto model at finite temperature and 
finite baryon number chemical potential. The chemical potential is 
introduced as an asymptotic value of the U(1) gauge field on 
the probe D8-$\overline{\text{D8}}$-branes. Using this model 
we studied the phase structure of the chiral symmetry breaking 
and obtained the phase diagram in Fig.~\ref{phase}. 
This phase diagram should be compared with that expected 
in QCD \cite{Kogut:2004su}. 
Our result is especially different from the QCD expectation 
at low temperature. 
In QCD the chiral symmetry is expected to be unbroken even at 
zero temperature if the chemical potential is sufficiently large. 
In our analysis the chiral symmetry is always broken 
below $T_c$ since the geometry of the $\tau$-$U$ space allows 
only the curved configuration of the probe branes. 
Since we used the probe approximation for the 
D8-$\overline{\text{D8}}$-branes, the gauge field on the branes 
(the chemical potential) does not affect the bulk geometry. 
It is interesting to see whether back-reactions of the gauge field 
on the geometry of the $\tau$-$U$ space change the phase structure 
below the temperature $T_c$. 
To fully understand the phase diagram we need an analysis 
beyond the probe approximation. 
\par
In the usual field theoretical approaches to the chiral symmetry 
breaking one considers condensations of quark bilinears 
$\bar{\psi}\psi$ as an order parameter. 
The quark masses are sources of these operators. 
In fact, in other models of the holographic QCD 
\cite{Babington:2003vm,Kruczenski:2003uq} the mechanism of the chiral 
symmetry breaking is different from that in the Sakai-Sugimoto model. 
The chiral symmetry is realized as the rotational symmetry of the 
probe branes in the transverse space. 
The quark masses and the quark condensations can be read from 
asymptotic behaviors of the probe branes. 
In the Sakai-Sugimoto model quarks are always massless since the 
asymptotic distance between the D4-branes and the 
D8-$\overline{\text{D8}}$-branes, 
which is proportional to the quark mass, is zero. 
It is not clear how to introduce quark masses in this model. 
It is interesting to clarify the relation between the mechanisms 
of the chiral symmetry breaking in the Sakai-Sugimoto model and 
in other models. 
\par
%
\vspace{10mm}
\noindent {\Large \textbf{Acknowledgements}}
\vspace{3mm}

We would like to thank T. Sakai and Y. Takada for useful discussions. 
One of us (N.H.) is grateful to T. Shimomura for discussions on 
numerical calculations. 

\bigskip

%

%

\end{document}